\documentclass[10pt,letterpaper]{article}
\usepackage[top=0.85in,left=2.75in,footskip=0.75in,marginparwidth=2in]{geometry}

\usepackage[utf8]{inputenc}

\usepackage{lastpage,fancyhdr,graphicx}
\usepackage{epstopdf}
\usepackage{cite} 
\usepackage{multirow}
\usepackage{url}
\usepackage{tabularx}
\usepackage{placeins}
\usepackage{multicol,lipsum}
\usepackage{rotating}
\usepackage{natbib}
\usepackage{cite}
\usepackage{amsmath}
\usepackage{url}
\usepackage{xcolor}
\definecolor{newcolor}{rgb}{.8,.349,.1}
\usepackage{mathtools, nccmath}
\usepackage{multicol, blindtext}
\usepackage{graphicx}
\usepackage{mathrsfs}
\usepackage{rotating}
\usepackage{blindtext}
\usepackage{mathptmx}
\usepackage{anyfontsize}
\usepackage{t1enc}
\usepackage{amsmath, amssymb, txfonts}
\usepackage{amsmath}
\usepackage{rotating}
\usepackage{lipsum}
\usepackage{natbib}
\usepackage{lipsum}
\usepackage{multicol}

\usepackage{lmodern}
\usepackage{amssymb}
\usepackage{latexsym}
\usepackage{mathtools}
\usepackage{url}
\usepackage{xcolor}
\usepackage{graphicx}
\usepackage{multicol}

\usepackage{nameref,hyperref}
\usepackage[right]{lineno}

\usepackage{microtype}
\DisableLigatures[f]{encoding = *, family = * }

\raggedright
\usepackage{changepage}
\usepackage[aboveskip=1pt,labelfont=bf,labelsep=period,singlelinecheck=off]{caption}

\makeatletter
\renewcommand{\@biblabel}[1]{\quad#1.}
\makeatother

\usepackage{lastpage,fancyhdr,graphicx}
\usepackage{epstopdf}
\usepackage{cite} 
\usepackage{multirow}
\usepackage{url}
\usepackage{tabularx}
\usepackage{multicol,lipsum}
\usepackage{rotating}
\usepackage{natbib}
\fancyheadoffset[L]{2.25in}
\fancyfootoffset[L]{2.25in}

\usepackage{color}

\definecolor{Gray}{gray}{.25}

\usepackage{graphicx}

\usepackage{sidecap}

\usepackage{wrapfig}
\usepackage[pscoord]{eso-pic}
\usepackage[fulladjust]{marginnote}
\reversemarginpar

\begin{document}
\vspace*{0.35in}

% title goes here:
\begin{flushleft}
{\Large
\textbf\newline{Characteristics of EUV Solar Coronal Bright Points Using Time Series Analyses}
}
\newline
% authors go here:
\\
Mahdieh Mehrabian\textsuperscript{1},
Bardia Kaki\textsuperscript{1, *},
Yusefali Abedini\textsuperscript{1}

\bigskip
\bf{1} {Department of Physics, University of Zanjan, University Blvd., 45371-38791 Zanjan, Iran}
\bigskip

*bardia.kaki@znu.ac.ir

\end{flushleft}
\justify
\section*{Abstract}

Coronal bright points (CBPs) as characteristics of the solar
corona are small-scale bright features ubiquitously observed at
extreme ultraviolet passbands. Here, we focused on time series
of CBPs in the periods of their lifetimes. We used Solar
Dynamic Observatory (SDO) / Atmospheric Imaging Assembly
(AIA) taken at 171 and 193 \AA. Using image processing
methods, CBPs were extracted from data and tracked during
their lifetimes. We found that CBPs observed in 171 \AA have
more lifetimes than those of observed at 193 \AA. CBPs at 171 \AA
are more fluctuated in intensity and appeared as highly density
number (DN) features than 193 \AA CBPs. It was found that
about $75 \%$ of CBPs are firstly achieved their peak at 171 \AA,
and then, after 12 seconds two minutes are achieved to their
peak at 193 \AA. Computing Pearson correlation for time series of
CBPs at both wavelengths where the correlation reaches at
maximum value gives information about time delay of magnetic
reconnection about 40 seconds from one layer to the other one.

\textbf{Key words:} Sun: corona, Sun: ultraviolet radiation, Sun: active
region (AR), technique: image processing

% the * after section prevents numbering
\section*{Introduction}

Solar coronal bright points (CBPs) are observable small-scale
features in X-ray and ultraviolet radiations (e.g., \cite{Vaiana_1973}; \cite{Habbal_1981}). \cite{Habbal_1981}
reported results of an analysis of Skylab observations about
coronal bright points made in EUV spectral lines formed in the
chromosphere, transition region and corona. It was found that
the diameter of CBPs is generally smaller than 60 arcsec 2
(\cite{Vaiana_1973}; \cite{Zhang2001}) and about $95\%$ of CBPs have
life times less than 20 hours. Their lifetimes take an extended range
from less than 20 minutes to more than 72 hours \citep{Alipour_2015}. \cite{Sheeley_1979} concluded that a CBP would
be consisted of two or three miniature loops (each $\sim 2500~km$ in diameter and $\sim 12000$~km long). 

On the other hand, The nature of internal structure of the coronal loops influence the heating mechanism. The properties of magnetic flux tubes and coronal loops oscillations have been studied by \cite{Esmaeili_2015, Esmaeili_2016}. Another interesting work as emphasis of internal structures about the importance of the interaction of multi-starnded coronal loops  have been investigated by \cite{Esmaeili_2016, Esmaeili_2017}. Most of the CBPs also play an important
role in the local heating of the plasma, and then, causes to lead
to magnetic reconnection \citep{Priest}.
Identifying and tracking CBPs derived from various solar data
are the two key procedures used to understand their physical
and statistical properties.  Canceling of magnetic bipoles may cause the majority of CBPs, also the emergence of magnetic flux are related to the few percent of CBPc \citep{Parnell, Priest_1994, Webb_1993, Zhang_2012}.

An automatic detection of bright
points in Yohkoh/SXT images was developed by \cite{Nakakubo_2000}. Also, an automatic detecting and tracking CBPs
using background intensity thresholds was proposed by \cite{McAteer_2005}. Then, an automatic algorithm for the
identification of CBPs in SOHO/EIT 193 \AA images was
investigated by \cite{Sattarov_2010}. \cite{Alipour_2015}
developed an automatic detection for CBPs at 193 \AA SDO/AIA
images based on feeding invariant image moments to the
support vector machine (proposed by \cite{Javaherian_2014}).
An extended overview of solar image processing techniques
employed for supervised and unsupervised automated feature
detection of small and large events can be found in
\cite{Aschwanden_2010}, \cite{Arish_2016}, and \cite{Javaherian_2017}.

Here, we employed SDO/AIA images at two wavelengths (193
\AA and 171 \aa) to investigate the variations of CBPs time series
occurred at both passbands simultaneously. Some statistical
properties (such as skewness and kurtosis) and the delay
between time series are studied based on Pearson correlation.

This paper is organized is as follows: the description of
datasets is introduced in section Data Analysis. The results
are given in section Results and Discussions. Finally,
concluding remarks are explained in section Conclusions.

\section*{Methods}
\begin{figure*}[ht]
\centerline{\includegraphics[width=1\textwidth,clip=]{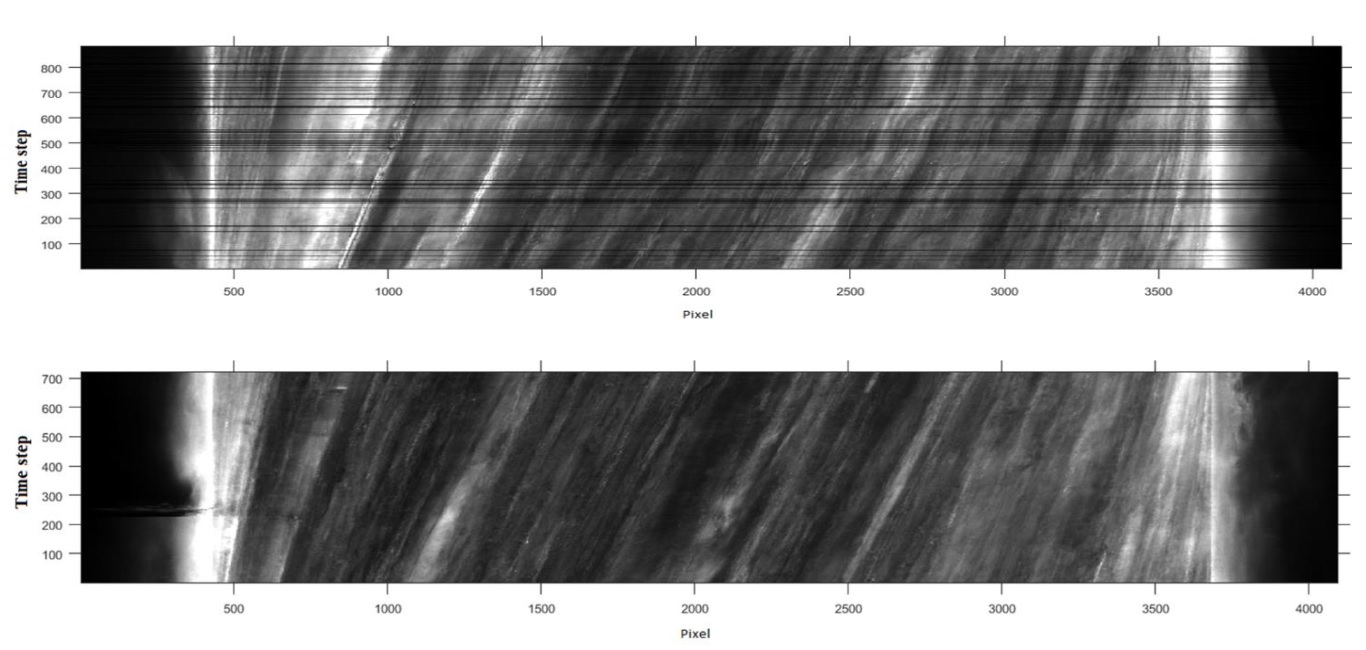}}
\caption{Making a spatial-time series from the average of images before removing the defective images (up) and after removing defective images (down)}\label{fig1}      
\end{figure*}

In this research, we wanted to look at the corona bright points. To this end, we started with collecting data on the corona bright points. The data are available at Stanford University's database\footnote{http://jsoc.stanford.edu.}. We received the data in the form of images at the wavelengths of 171 and 193 \AA with Fits suffix. Due to the current limits, the time interval of each image to the next one is 12 seconds. Therefore, by developing time series of these images, we expect to have a spot for every 12 seconds. However, sometimes some images may be flawed or cracked due to various reasons, which is a familiar thing in the field of imaging. In the present research, it has been tried to select only those coronal bright points without flaws and cracks in their time series as much as possible to form a regular and perfect time series. This is more possible for the formation of time series of bright points that occur at a time other than the time of the flare. Nevertheless, the presence of an empty gap and the absence of one or more consecutive images at the time of the flare are inevitable; since at this time, the light changes and the rate of radiation entering the lens of the camera are so much that they creates inappropriate images for data analysis. In Fig. \ref{fig1}, these defective images can be seen as dark lines beside the healthy images. The way of creating \ref{fig1} is that, first three rows of the middle pixels of each image are chosen, and then averaging is performed on these three rows. Then, this average is considered as the representative of that image, being given a row of the new image. The same process is also repeated for the next image to be placed as the next row in the image on the prior row. Ultimately, the new image includes a series of averaged images aligned to show us the spatial-time changes of the sun in the studied interval. Using this method, images with much more or less light intensity than their previous and next images can be regarded as the defective images.

Automatic programs can be used to measure this; however, in the present study, this measurement has been done with eyes. As seen in Figs. , by placing consecutive images on each other, something like the visualization of a whirlwind can be imagined that is due to the sun's differential spin. The sun rotates around its own axis, making the points on the sun move from their original location in a received series of times. Certainly, this spin must be removed to examine a region on the sun, but the elimination of this spin is a little complicated, since the entire surface of the sun does not move at the same rate. The sun's rotation can be seen in the latitudes close to the equator of the sun, once every 25 days, and for the latitudes close to the poles once every 36 days. This type of rotation, called the differential rotation, causes a phenomenon named solar dynamo.

\begin{figure*}[ht]
\centerline{\includegraphics[width=1.05\textwidth,clip=]{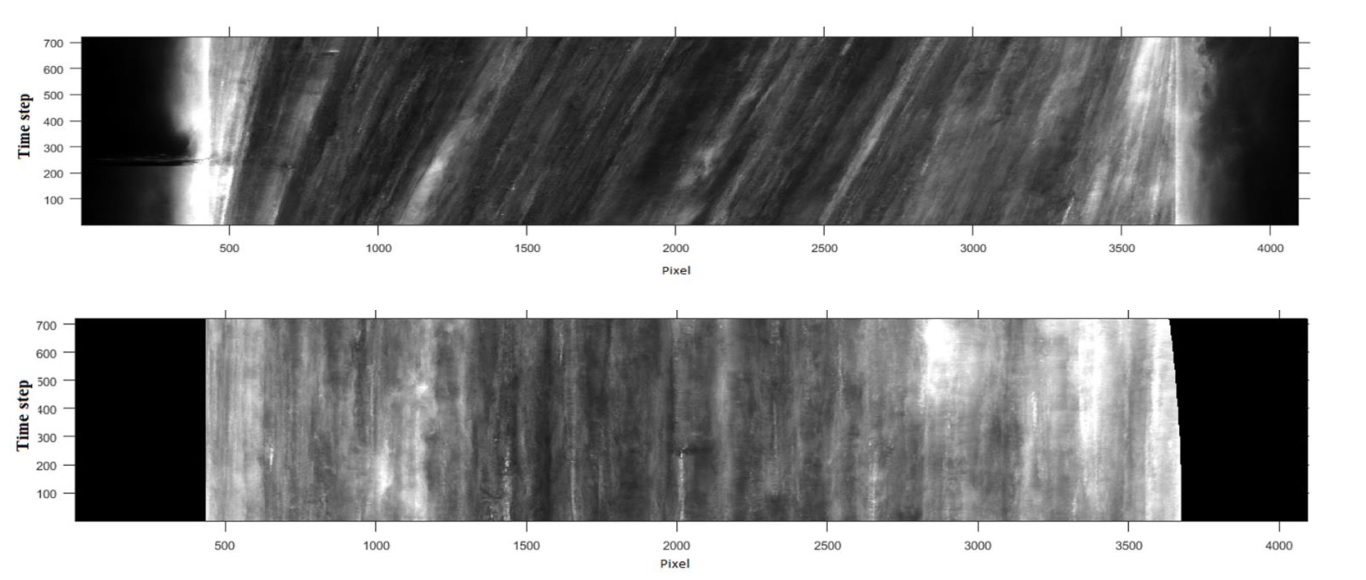}}
\caption{Removing differential rotation.}\label{fig2}      
\end{figure*}

\begin{figure*}[ht]
\centerline{\includegraphics[width=.9\textwidth,clip=]{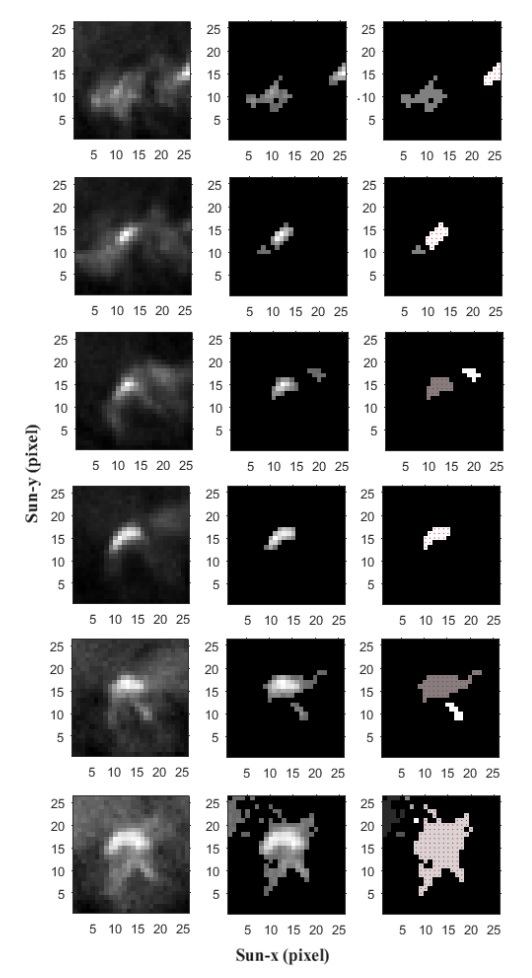}}
\caption{Transformation of a bright points in the 171 \AA wavelength channel.}\label{fig3}.     
\end{figure*}

\begin{figure*}[ht]
\centerline{\includegraphics[width=.9\textwidth,clip=]{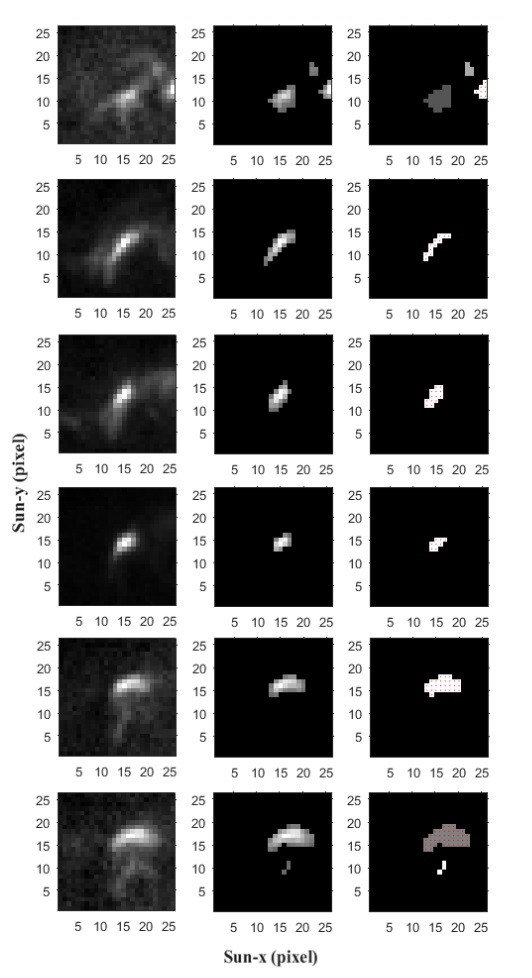}}
\caption{Transformation of a bright points in the 193 \AA wavelength channel.}\label{fig4}.     
\end{figure*}

In the current study, this effect has been eliminated using the ssw package in the IDL environment. In Fig. \ref{fig2}, an example of the removal of the differential rotation is indicated. After initial processing and data preparation, for further investigation, we can now build time series with these data.  We separated the corona bright points at the wavelengths of 171 and 193 \AA. In Figs. \ref{fig3} and \ref{fig4}, the changes in a corona bright spot are shown in both wavelengths of 171 and 193 \AA. As shown in the figures, changes in a bright spot represent a different behavior in two wavelengths; and the area occupied by the bright spot can have different environments and areas. In these figures, the boundaries were first separated by defining the light threshold intensity of the received pixels. Then, using the region growth method described, we separated the points in the boundaries. In the middle image of the Figs. \ref{fig3} and \ref{fig4}, this separation of the bright spot from the background is observable. In a given frame, since there may be several regions with a light intensity greater than the threshold, the region with the maximum intensity was considered as the coronal bright points.

This method was applied to the software and its visual representation can be observed in the form of small red dots among the brightest pixels specified in the right parts of both images 3-5 and 4-5. Each row shown in Figs. 3-5 and 4-5 has been selected on a regular basis in order to display, so that each row has a difference of 10 images from the row after itself. In addition, as mentioned in Chapter 3, the time interval of each image to the next image on SDO satellite is twelve seconds; thus, each row of Figs. \ref{fig3} and \ref{fig4} is about 120 seconds different from the next row. For this specific bright points shown in Figs. \ref{fig3} and \ref{fig4}, the total length of the time series is 600 seconds or about ten minutes. The coronal bright points have lifetime of about a few minutes to a few tens of minutes. Nevertheless, since one of the purposes of this study was to also investigate the corona bright points at the occurring time of the flare and not occurring time of the flares, we were forced to choose bright points comparable to the lifetimes of a solar flare. Thus, the bright points studied in this research have lifetime of 10 to 30 minutes. As mentioned, the lifetime of the coronal bright points is directly proportional to their sizes, i.e. the larger the bright spot is, the longer it will live, and vice versa.

\section*{Summary and result}
\begin{figure*}[ht]
\centerline{\includegraphics[width=.9\textwidth,clip=]{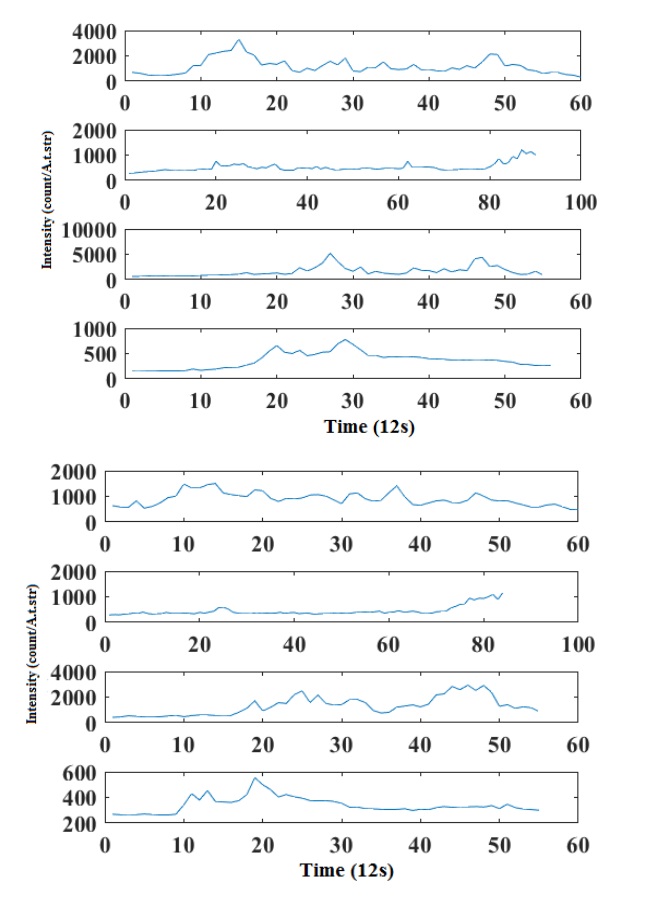}}
\caption{Four time-series examples of corona bright points at 171 \AA (up) and 193 \AA wavelengths before the occurrence of flares.}\label{fig5}     
\end{figure*}

\begin{figure*}[ht]
\centerline{\includegraphics[width=.9\textwidth,clip=]{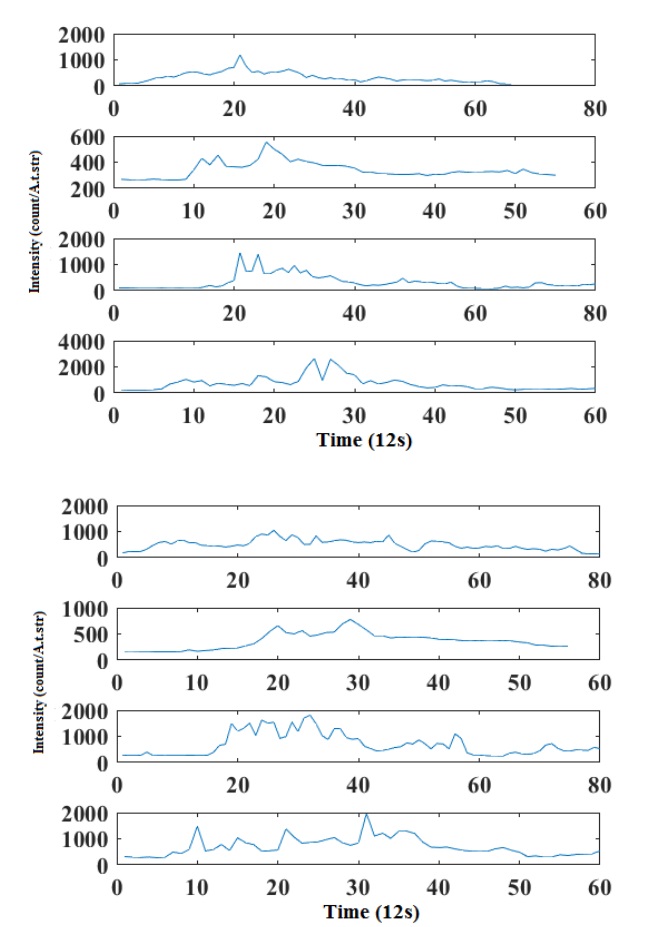}}
\caption{Four time-series examples of corona bright points at 171 \AA (up) and 193 \AA wavelengths at the occurrence of flares.}\label{fig6}  
\end{figure*}

\begin{figure*}[ht]
\centerline{\includegraphics[width=.9\textwidth,clip=]{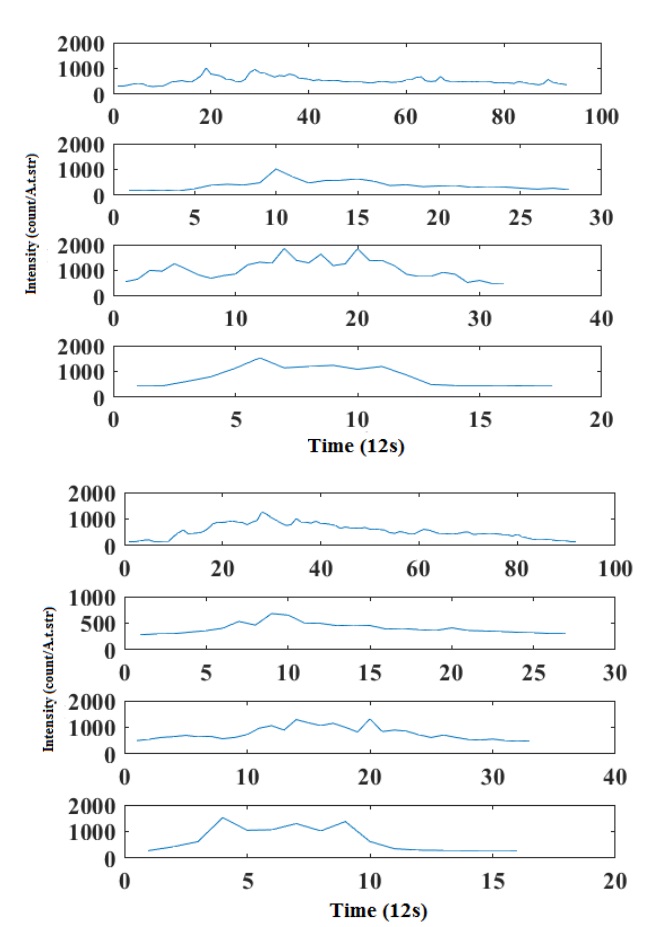}}
\caption{Four time-series examples of corona bright points at 171 \AA (up) and 193 \AA wavelengths after the occurrence of flares }\label{fig7}  
\end{figure*}

Accordingly, the research goals led us to focus on the coronal bright points with the lifetime of less than 30 minutes. In the current study, 24 coronal bright points have been separately studied at wavelengths of 171 and 193 \AA, before occurrence of solar flares, during the solar flares and after the solar flares. It has been tried to select bright points from different areas of the corona as much as possible. All of the studied bright points in this study are for the date 2011-02-13, when a flare with the scale of 6.6 Mega meter has occurred. The time series graph of the coronal bright points studied in this research is represented in Figs. \ref{fig5}-\ref{fig7}. By the help of these time series, we can extract many statistical moments used in processes such as simulation of corona bright points.

We are looking for parameters such as minimum, maximum, mean, standard deviation, skewness, and kurtosis for 24 time series in the 171 and 193 \AA wavelength channels.
According to the results obtained from this study, the minimum brightness of a corona bright spot in the 171 \AA channel is 170.32 $(DN/A.t.str)$ , where DN is the photon numerical density, A stands for area, t represents time, and str shows the angular Arc. The minimum brightness in the 193-angstroms wavelength channel is 84.79 with the mentioned unit. The maximum brightness of the 171 \AA wavelength channel is 5407.5 $(DN/A.t.str)$ and in the wavelength channel of 193 \AA, it is 2393 $(DN/A.t.str)$. These values provide us with the high and low brightness levels in the wavelengths of 171 and 193 angstroms, which can be useful tools for simulating the time series of corona bright points. The mean of the total intensity obtained in the 171 \AA wavelength channel is 807.12 and in the 193 \AA wavelength channel, it is 662.04 with the same unit as above.

Equation \ref{eq1} is the time series variance of any coronal bright spot at the desired wavelength.
\begin{equation}\label{eq1}
    variance = \frac{1}{N-1}\sum_{j=1}^{N}(A_{j}-\mu)^2
\end{equation}
In this equation, $A_{j}$ is any spot in the time series, which is calculated from one to the size of the lifetime of the coronal bright spot and $\mu$ is considered the mean of each time series. The upper limit of this parameter in the 171-angstroms wavelength channel is 980114.10 and its lower limit in this wavelength channel is 4721.36. Moreover, in the 193 \AA wavelength channel, the upper limit is 529211.22 and its lower limit is 2129.07 for the studied time series. 

The other statistical quantity in this study is skewness shown in Equation \ref{eq2}
\begin{equation}\label{eq2}
    skewness = \frac{1}{N}\sum_{j=1}^{N}(A_{j}-\mu)^3/\left(\sqrt{\frac{1}{N}\sum_{j=1}^{N}(A_{j}-\mu)^2}\right)
\end{equation}

Given the positive skewness in table ((\ref{tab1}-\ref{tab6})), it can be claimed that all the brightness, like the total brightness of the sun, has a sequence to the right that confirms the point that the brightness distribution function of the bright points also follows the log-normal function.

There is another quantity called kurtosis of equation \ref{eq3}, indicating the peak of a probable distribution (or deviation from the normal distribution). 
\begin{equation}\label{eq3}
    kurtois = \frac{1}{N}\sum_{j=1}^{N}(A_{j}-\mu)^4/\left(\frac{1}{N}\sum_{j=1}^{N}(A_{j}-\mu)^2\right)^2
\end{equation}
The mean for this quantity for the 171 \AA wavelength channel is 3.92 and for the 193 \AA wavelength channel, it is 4.92 According the length of the time series and based on what obtained from the mean kurtoses at the two wavelengths of 171 and 193 \AA, it can be said that at the wavelength of 193\AA with more kurtosis, the number of brightness with the same scale is greater than the observed value at the wavelength of 171 \AA. All values mentioned in Table ((\ref{tab1}-\ref{tab6})) have been represented for each time series.

Another part of this research is dedicated to finding the relationship between corona bright points at different wavelengths. This relationship is obtained by the help of correlation between the time series of a bright spot at the two wavelengths of 171 and 193 \AA. The Pearson correlation is the most famous and common form of correlation defined as follows:
\begin{equation}\label{eq4}
correlation (A,B) =\left(\sum_{j=1}^{N}(A_{j}-\mu_{A})(B_{j}-\mu_{B})\right)/\sqrt{\sum_{j=1}^{N}(A_{j}-\mu_{A})^2\sum_{j=1}^{N}(B_{j}-\mu_{B})^2}
\end{equation}
The correlations represented in Tables (\ref{tab1}-\ref{tab6}) that have values of about 0.5 or higher indicate the strong relationship between the wavelengths of 171 and 193 \AA of a coronaL right spot. This indicates that the different layers of the corona are related to each other and affect each other. The significant point in this regard is that this relationship is associated with a delay in most cases. That is, it takes a while for an event in a layer, like 171 \AA, to put its effect on layer 193 \AA. The upper limit of this time delay for the studied bright points was 2.5 minutes and its lower limit was 12 seconds, which is likely to be influenced by the limit of the instruments.
When the data is statistically studies, we need to be able to prove that the significant relationship between the two statistical sets, if any, is not random. An appropriate method to prove this is the method of probability value $(p value)$ test. We can use the contradiction method for this purpose, so that first a value $H_{0}$ is considered as a null hypothesis value, then if the obtained value is equal to H1, if $H_{0} \neq H_{1}$, this hypothesis is rejected and the relationship is significant. In order to define a null set, we need a threshold $(\alpha)$. This threshold is different in various sciences. For instance, in the humanities, this threshold is set at 0.05 and in the medical sciences at 0.01. In some sources, this threshold has also been introduced as the effective level. If the obtained p value is less than $(\alpha)$, then the null hypothesis is rejected and the proposed hypothesis is confirmed.

The how to calculate the probability test criterion (K) is shown in Equation \ref{eq5}
\begin{equation}\label{eq5}
K=\frac{\bar{x}-\mu}{\frac{\delta}{\sqrt{n}}}
\end{equation}
The region where the probability of K test criteria is equal to or smaller than the probability of $(\alpha)$ is called a critical region. Since the probability level is the level where the event does not occur, the critical region is the numerical region between $\mu - 3\delta$ and $\mu + 3\delta$. Finally, the primary hypothesis can be confirmed or rejected by comparing the numerical value obtained for the test criterion and the critical regions. The value of this test in the current study is from the level of $10^{-23}$ to $10^{-21}$, which is much less than the threshold of 0.05 or 0.01, indicating that the obtained correlation is fully valid and significant.

\begin{figure*}[ht]
\centerline{\includegraphics[width=.9\textwidth,clip=]{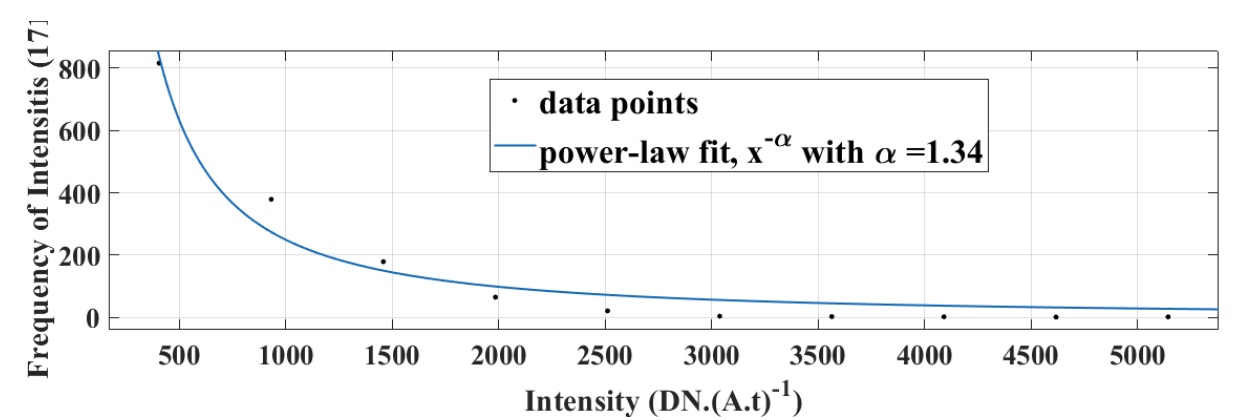}}
\caption{Average distribution function of the intensities of corona bright points points at the wavelength of 171 \AA.}\label{fig8}
\end{figure*}

\begin{figure*}[ht]
\centerline{\includegraphics[width=.9\textwidth,clip=]{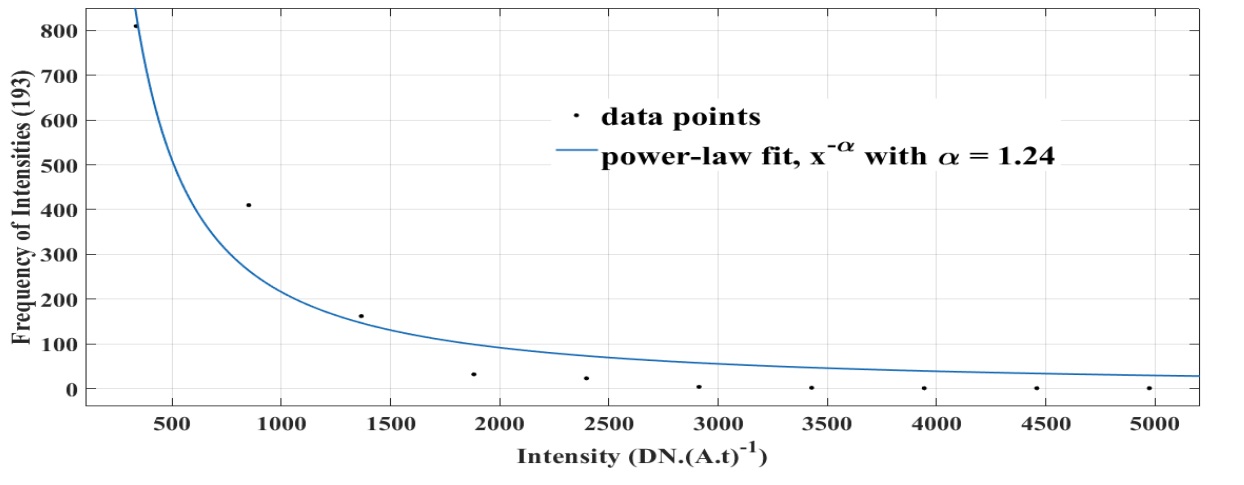}}
\caption{Average distribution function of the intensities of corona bright points points at the wavelength of 193 \AA.}\label{fig9}  
\end{figure*}

The mean distribution functions of the intensity of the corona bright points for all points of the received data at the wavelengths of 171 and 193 angstroms are respectively represented in Figs. \ref{fig8} and \ref{fig9}. As observed in these images, the intensity distribution functions follow a power function as $y \propto x^{\alpha}$. In these figures, the fitted slopes $(\alpha)$ have been obtained $-1.34$ and $-1.24$ at a wavelength of 171 and 193 angstrom, respectively. The fit has been performed by the least squares method, obtained 0.96 and 0.94 for the wavelengths of 171 and 193 angstroms, respectively.

\begin{sidewaystable}[]

\caption{Statistical momentum with correlation for the studied bright points (events: 1-4)}
\label{tab1}

\begin{tabular}{|l|l|l|l|l|l|l|l|l|}
\hline
\multirow{2}{*}{}Number of event & \multicolumn{2}{l|}{1} & \multicolumn{2}{l|}{2} & \multicolumn{2}{l|}{3} & \multicolumn{2}{l|}{4} \\ \cline{2-9} 
                  &      171     &         193  &      171     &193           &      171     & 193          &171           &    193       \\ \hline
           Length of time series (12 sec)       &   85        &79           &   50        &49           &98           &105           &  74         &103           \\ \hline
                  Minimum& 388.148          & 255.15          & 531.47          &       346.59    &    257.94       & 201.59          &234.6           &185.3           \\ \hline
              Maximum    &1957.6           &    1409.33       &2339.8           &1903.25           &1921.75           &1822.7           &497.07           &         512.44  \\ \hline
                 Mean & 956.32          &  586.24         &1282.41           & 902.06          &674.61           &      694.41     &    323.44       & 236.98          \\ \hline
                Variance  &      184061.32     &84414.24           &263906.41           &  145449.74         &121183.66           &        165264.88   &3721.36           & 2129.07          \\ \hline
                Skewness  & 0.7          & 1.42          & 0.0012          &0.21           &    1.24       &1.09           &0.52           &    2.97       \\ \hline
                Kurtosis  &      2.45     & 4.17          &1.94           & 2.39          & 3.75          &     3.34      &      2.43     &15.73           \\ \hline
                  Pearson correlation& \multicolumn{2}{l|}{0.84} & \multicolumn{2}{l|}{0.71} & \multicolumn{2}{l|}{0.87} & \multicolumn{2}{l|}{0.62} \\ \hline
                  $P$ value& \multicolumn{2}{l|}{0.0} & \multicolumn{2}{l|}{0.0} & \multicolumn{2}{l|}{0.0} & \multicolumn{2}{l|}{0.0} \\ \hline
                Delay (12 sec)  & \multicolumn{2}{l|}{4} & \multicolumn{2}{l|}{1} & \multicolumn{2}{l|}{4} & \multicolumn{2}{l|}{10} \\ \hline
                
\end{tabular}

\end{sidewaystable}

\begin{sidewaystable}[]
\caption{Events: 5-8}
\label{tab2}
\begin{tabular}{|c|c|c|c|c|c|c|c|c|}
\hline
\multirow{2}{*}{}Number of event & \multicolumn{2}{l|}{1} & \multicolumn{2}{l|}{2} & \multicolumn{2}{l|}{3} & \multicolumn{2}{l|}{4} \\ \cline{2-9} 
                  &      171     &         193  &      171     &193           &      171     & 193          &171           &    193       \\ \hline
           Length of time series (12 sec)       &   60        &60           &   90        &84           &55           &55           &  80         &66           \\ \hline
                  Minimum& 318.66         & 475.89          & 287.82          &       285.85    &    623.42       & 423.21          &141.98           &77.58           \\ \hline
              Maximum    &3318.75           &    1504.83       &1199.87           &1150.16           &5232.25           &2933           &1046           &         1191.40  \\ \hline
                 Mean & 1170.29          &  902.09         &518.53           & 447.13          &1631.05           &      1307.95     &    491.21       & 333.46          \\ \hline
                Variance  &    370290.42       &65687.89           &29743.43          &  36960.61         &989114.10           &529211.22  &38309.94           & 41449.20          \\ \hline
                Skewness  & 1.1         & 0.54         & 2.09         &  2.23           &    1.7       &0.65   &0.41           &    1.4       \\ \hline
                Kurtosis  &      4.35   & 2.69          &7.54          & 6.96          & 5.88          &     2.44     &      2.88     &6.14           \\ \hline
                  Pearson correlation& \multicolumn{2}{l|}{0.65} & \multicolumn{2}{l|}{0.8} & \multicolumn{2}{l|}{0.71} & \multicolumn{2}{l|}{0.63} \\ \hline
                  $P$ value& \multicolumn{2}{l|}{0.0} & \multicolumn{2}{l|}{0.0} & \multicolumn{2}{l|}{0.0} & \multicolumn{2}{l|}{0.0} \\ \hline
                Delay (12 sec)  & \multicolumn{2}{l|}{1} & \multicolumn{2}{l|}{6} & \multicolumn{2}{l|}{1} & \multicolumn{2}{l|}{4} \\ \hline
\end{tabular}
\end{sidewaystable}

\begin{sidewaystable}[]

\caption{Events: 9-12}
\label{tab3}
\begin{tabular}{|c|c|c|c|c|c|c|c|c|}
\hline
\multirow{2}{*}{}Number of event & \multicolumn{2}{l|}{1} & \multicolumn{2}{l|}{2} & \multicolumn{2}{l|}{3} & \multicolumn{2}{l|}{4} \\ \cline{2-9} 
                  &      171     &         193  &      171     &193           &      171     & 193          &171           &    193       \\ \hline
           Length of time series (12 sec)       &   56       &55           &   80        &80           &60           &60           &  70         &43           \\ \hline
                  Minimum& 158.41         & 363.58          & 250.09          &       79.84    &    275.5       & 220.61          &329.38           &412.12           \\ \hline
              Maximum    &777.12          &    555.2       &181.26           &1464.33           &1963.5           &3634      &2584.66          &         2240  \\ \hline
                 Mean & 367.69          &  342.14         &684.96           & 315.92          &696.7           &      712.83     &    712.14       & 770.84          \\ \hline
                Variance  &    24003.16       &3820.57           &185796.58          &  78155.64         &123169.92           &291933.69  &31522.89           & 236371.72          \\ \hline
                Skewness  & 0.46         & 1.1         & 0.96         &  2.01           &    1.16       &1.92   &1.94           &    1.55       \\ \hline
                Kurtosis  &      2.67  & 4.5          &2.83          & 7.61          & 4.46          &     6.77     &      5.8     &4.3           \\ \hline
                  Pearson correlation& \multicolumn{2}{l|}{0.48} & \multicolumn{2}{l|}{0.8} & \multicolumn{2}{l|}{0.41} & \multicolumn{2}{l|}{0.86} \\ \hline
                  $P$ value& \multicolumn{2}{l|}{0.0} & \multicolumn{2}{l|}{0.0} & \multicolumn{2}{l|}{0.0} & \multicolumn{2}{l|}{0.0} \\ \hline
                Delay (12 sec)  & \multicolumn{2}{l|}{1} & \multicolumn{2}{l|}{2} & \multicolumn{2}{l|}{2} & \multicolumn{2}{l|}{9} \\ \hline
\end{tabular}
\end{sidewaystable}

\begin{sidewaystable}[]

\caption{Events: 13-16}
\label{tab4}
\begin{tabular}{|c|c|c|c|c|c|c|c|c|}
\hline
\multirow{2}{*}{}Number of event & \multicolumn{2}{l|}{1} & \multicolumn{2}{l|}{2} & \multicolumn{2}{l|}{3} & \multicolumn{2}{l|}{4} \\ \cline{2-9} 
                  &      171     &         193  &      171     &193           &      171     & 193          &171           &    193       \\ \hline
           Length of time series (12 sec)       &   60   &39           &   90        &61           &62           &60           &  54         &54           \\ \hline
                  Minimum& 254.02       & 395.46          & 289.13          &       343.73    &    353.63       & 583.86          &489.07           &550.19           \\ \hline
              Maximum    &1787.57         &    1323.66       &5407.50           &2425.50           &2249.00           &2005.80      &2500.40          &         2640.25  \\ \hline
                 Mean & 737.01          &  657.84         &1164.60           & 707.00          &971.47           &      1060.07     &    1123.33       & 1119.35         \\ \hline
                Variance  &    24475.68       &81866.62           &678096.60          &  106101.24         &161021.71           &194938.11  &303968.22           & 275818.26          \\ \hline
                Skewness  & 0.66         & 0.79         & 2.52         &  2.97           &    0.6       &0.45   &0.74           &    1.2       \\ \hline
                Kurtosis  &      1.96  & 2.19          &11.7          & 14.7      & 3.38          &     1.83     &      2.6     &3.96           \\ \hline
                  Pearson correlation& \multicolumn{2}{l|}{0.66} & \multicolumn{2}{l|}{0.68} & \multicolumn{2}{l|}{0.78} & \multicolumn{2}{l|}{0.84} \\ \hline
                  $P$ value& \multicolumn{2}{l|}{0.0} & \multicolumn{2}{l|}{0.0} & \multicolumn{2}{l|}{0.0} & \multicolumn{2}{l|}{0.0} \\ \hline
                Delay (12 sec)  & \multicolumn{2}{l|}{10} & \multicolumn{2}{l|}{14} & \multicolumn{2}{l|}{4} & \multicolumn{2}{l|}{1} \\ \hline
\end{tabular}
\end{sidewaystable}

\begin{sidewaystable}[]
\caption{Events: 17-20}
\label{tab5}
\begin{tabular}{|c|c|c|c|c|c|c|c|c|}
\hline
\multirow{2}{*}{}Number of event & \multicolumn{2}{l|}{1} & \multicolumn{2}{l|}{2} & \multicolumn{2}{l|}{3} & \multicolumn{2}{l|}{4} \\ \cline{2-9} 
                  &      171     &         193  &      171     &193           &      171     & 193          &171           &    193       \\ \hline
           Length of time series (12 sec)       &   18   &16           &   35        &35       &93           &92           &  28         &27           \\ \hline
                  Minimum& 433.86       & 286.71          & 232.17          &       105.02    &    296.03       & 148.02          &170.32           &282.82           \\ \hline
              Maximum    &1527.40         &    1522.25       &1178.00           &978.87           &1022.50           &1269.05      &1021.00          &         681.12  \\ \hline
                 Mean & 799.36         &  694.59         &669.41           & 386.16          &530.37           &     549.55     &    395.81       & 405.88         \\ \hline
                Variance  &    138136.77      &203604.18          &74286.27          &  66441.85         &20099.29           &72739.56  &35747.63           & 10467.51         \\ \hline
                Skewness  & 0.4         & 0.58         & 0.11         &  1.1           &    1.08      &0.3   &1.41          &    1.16       \\ \hline
                Kurtosis  &      1.69  & 1.78          &1.98          & 3.11      & 4.31          &     2.36     &      5.43     &3.9           \\ \hline
                  Pearson correlation& \multicolumn{2}{l|}{0.82} & \multicolumn{2}{l|}{0.68} & \multicolumn{2}{l|}{0.8} & \multicolumn{2}{l|}{0.79} \\ \hline
                 $P$ value & \multicolumn{2}{l|}{0.0} & \multicolumn{2}{l|}{0.0} & \multicolumn{2}{l|}{0.0} & \multicolumn{2}{l|}{0.0} \\ \hline
                Delay (12 sec)  & \multicolumn{2}{l|}{3} & \multicolumn{2}{l|}{1} & \multicolumn{2}{l|}{1} & \multicolumn{2}{l|}{2} \\ \hline
\end{tabular}
\end{sidewaystable}

\begin{sidewaystable}[]
\caption{Events: 21-24}
\label{tab6}
\begin{tabular}{|c|c|c|c|c|c|c|c|c|}
\hline
\multirow{2}{*}{}Number of event & \multicolumn{2}{l|}{1} & \multicolumn{2}{l|}{2} & \multicolumn{2}{l|}{3} & \multicolumn{2}{l|}{4} \\ \cline{2-9} 
                  &      171     &         193  &      171     &193           &      171     & 193          &171           &    193       \\ \hline
           Length of time series (12 sec)       &   33   &32           &   45        &38       &81           &77           &  28         &25           \\ \hline
                  Minimum& 497.67       & 477.99          & 242.97         &       125.02    &    216.47       & 167.41          &150.32           &232.72           \\ \hline
              Maximum    &1854.75        &    1321.25       &978.00           &778.87           &1129.20           &969.73      &1229.14          &         883.40  \\ \hline
                 Mean & 1039.95         &  770.33         &463.41           & 216.86          &580.11           &     479.55     &    475.52       & 505.28         \\ \hline
                Variance  &    137724.21     &58892.99          &73127.27          & 63151.85         &21523.29           &68159.56  &34189.63           & 11431.51         \\ \hline
                Skewness  & 0.44         & 0.78         & 1.11         &  1.1           &    1.08      &0.3   &1.41          &    1.16       \\ \hline
                Kurtosis  &      2.49  & 2.52          &1.58          & 3.11      & 1.31          &     2.36     &      1.43     &2.2          \\ \hline
                  Pearson correlation& \multicolumn{2}{l|}{0.92} & \multicolumn{2}{l|}{0.62} & \multicolumn{2}{l|}{0.8} & \multicolumn{2}{l|}{0.79} \\ \hline
                  $P$ value & \multicolumn{2}{l|}{0.0} & \multicolumn{2}{l|}{0.0} & \multicolumn{2}{l|}{0.0} & \multicolumn{2}{l|}{0.0} \\ \hline
                Delay (12 sec)  & \multicolumn{2}{l|}{1} & \multicolumn{2}{l|}{1} & \multicolumn{2}{l|}{3} & \multicolumn{2}{l|}{2} \\ \hline
\end{tabular}
\end{sidewaystable}

\section*{Conclusion}
Since the power distribution function has sequences to the right, it is true to expect that the skewness obtained in Tables (\ref{tab1}-\ref{tab6} for the coronal bright points has a positive value (a sequence to the right). The greater skewness of the distribution function obtained at 171 \AA wavelength indicates the minimum mean of intensities in this wavelength compared to the 193\AA wavelength.
As we know, on the surface of the sun, the energy of flares follows the power rule. Since the intensity of the corona right points also follows the same rule, and on the other hand, the energy is a function of intensity, it may be pointed out that like flares, the corona bright points have an explosive mechanism and follow the same rules. Since many methods are available for the simulation of flares, it is believed that by enhancing observations and extracting more statistical parameters, it is possible to build similar simulations for corona bright points. In addition, perhaps by finding more common features between corona bright points and solar flares, it would be possible to obtain a good approximation of the simulations of corona bright points by a slight change in the parameters and boundary conditions of the existing solar flares simulations.

\nolinenumbers

\FloatBarrier
%\bibliographystyle{plain}
%\setcitestyle{authoryear}
\bibliography{library}
\bibliographystyle{rusnat}

\end{document}